# Helium effects and bubbles formation in irradiated Ti$_3$SiC$_2$

Hongliang Zhang[1, 2, *, §], Ranran Su[1, 2, §], Izabela Szlufarska[2, *], Liqun Shi[1, *], Haiming Wen[3, *]

[1] Institute of Modern Physics, Fudan University, Shanghai, China

[2] Department of Materials Science and Engineering, University of Wisconsin, Madison, WI, 53706, USA

[3] Department of Materials Science and Engineering, Missouri University of Science and Technology, Rolla, U.S.A.

§ Co-first authors

* Corresponding authors.

zhlcanes@hotmail.com (H. Zhang)

szlufarska@wisc.edu (I. Szlufarska),

lqshi@fudan.edu.cn (L.Q. Shi),

wenha@mst.edu (H.M. Wen)

**Abstract**

Ti$_3$SiC$_2$ is a potential structural material for nuclear reactor applications. However, He irradiation effects in this material are not well understood, especially at high temperatures. Here, we compare the effects of He irradiation in Ti$_3$SiC$_2$ at room temperature (RT) and at 750 °C. Irradiation at 750 °C was found to lead to extremely elongated He bubbles that are concentrated in the nano-laminate layers of Ti$_3$SiC$_2$, whereas the overall crystal structure of the material remained intact. In contrast, at RT, the layered structure was significantly damaged and highly disordered after irradiation. Our study reveals that at elevated temperatures, the unique structure of Ti$_3$SiC$_2$ can accommodate large amounts of He atoms in the nano-laminate layer, without compromising the structural stability of the material. The structure and the mechanical tests results show that the irradiation induced swelling and hardening at 750 °C are much smaller than those at RT. These results indicate that Ti$_3$SiC$_2$ has an excellent resistance to accumulation of radiation-induced He impurities and that it has a considerable tolerance to irradiation-induced degradation of mechanical properties at high temperatures.

## 1. Introduction

Ti$_3$SiC$_2$, first synthesized in the 1960s [1], is one of the MAX phase materials, in which M is an early transition metal element, A is a group III or IV element, and X is C or N. Ti$_3$SiC$_2$ has many excellent properties, including good electrical and thermal conductivities ($11\times10^6$ $\Omega^{-1}m^{-1}$ and 43 W/m K, respectively) [2–5], high Young's modulus (~325 GPa) [6], high resistance to thermal shock, and good ductility (above 1100 °C) [7]. It has been previously reported, from both experiments and theory, that $Ti_3SiC_2$ exhibits a high resistance to irradiation-induced structural disordering [7-12]. Specifically, under ion irradiation, $Ti_3SiC_2$ was found to remain crystalline even at a very high dose of 116.9 displacements per atom (dpa) [11]. Furthermore, experiments have shown that at temperatures above 400 °C, $Ti_3SiC_2$ has a better resistance to radiation-induced cracking as compared with other MAX phase materials, such as $Ti_3AlC_2$ and $Ti_2AlC$ [9,12,13].

Thanks to the above properties, $Ti_3SiC_2$ could be potentially used as a structural material in Gen IV fission and fusion reactors where materials are exposed to high temperatures and intense neutron irradiation. For these applications, it is important to consider the behavior of He atoms, introduced either by radiation or by transmutation through the (n, α) nuclear reaction. In many materials, He atoms are known to interact with lattice vacancies, forming He–vacancy clusters, and eventually growing into He bubbles [14]. Once the He bubbles grow to a certain size (i.e., have a critical volume), they will rupture, causing the material to peel and flake off, which in turn will have a significant effect on mechanical properties of the material [15–17]. There have been several studies on the He irradiation of $Ti_3SiC_2$ [18–20], however, most of them are at RT. The only high-temperature study was at 450 °C, in which the irradiated $Ti_3SiC_2$ was analyzed by grazing incidence X-ray diffraction (GIXRD) [19]. The results showed smaller anisotropic swelling and less damage as compared to those at RT. However, the morphology and distribution of He bubbles were not studied in Ref. [19]. Some He irradiation studies at elevated temperatures have been reported for other MAX phases. For instance, the authors of Refs. [21–23] observed the morphology of He bubbles in He-irradiated $Ti_2AlC$ and $Ti_3AlC_2$ by bright-field transmission electron microscopy (TEM). However, due to the limitation of the resolution, the clear shapes of the He bubbles and their positions in the lattice could not be clearly identified. Elongated He bubbles were mentioned in Ref. [24], but the paper was focused on the radiation-induced phase transformation. In general, understanding of shapes and positions of He bubbles in MAX phases at elevated temperatures is still limited. In addition, the relationships between the morphology and distribution of He bubbles and the changes of the microstructure and in the mechanical properties at different temperatures have not been reported so far.

In this paper, we report a comparative study of high temperature and RT He irradiation in $Ti_3SiC_2$. Structural changes in irradiated $Ti_3SiC_2$ samples were analyzed with GIXRD and Rietveld refinements. The near-surface changes in the irradiated samples were characterized using Raman spectroscopy. Radiation-induced changes in mechanical properties (hardness and Young's modulus) were studied using nanoindentation. Changes in the microstructure and the He bubble morphology for samples irradiated at different temperatures were investigated using TEM. In particular, $Ti_3SiC_2$ structure consists of Ti-C layers interlaid with Si layers. The role of the layered structures in $Ti_3SiC_2$ in accommodating damage during He irradiation is also analyzed in this study using TEM and high resolution TEM.

## 2. Experimental Methods

The material used in this work was polycrystalline bulk $Ti_3SiC_2$, prepared by reactive sintering. Stoichiometric mixtures of $3Ti + SiC + C$ were prepared by hand grinding fine Ti (99.9%), SiC (99.9%), and C (graphite, 99.99%) powders under argon, followed by cold pressing in a hardened steel die at 180 MPa. The powders contained ~2 wt.% Al to assist with reactivity. The pressed cylindrical samples were sintered under flowing argon gas by heating to 1600°C at 10°C min$^{-1}$, holding for 4 h, and returning to RT. During sintering, a small amount of $Al_2O_3$ was formed in the sample. The as-sintered specimens were polished using fine metallographic abrasive paper and $Al_2O_3$ suspensions, cleaned by rinsing in ultrasonic baths of acetone and ethanol, and annealed at 800°C in a vacuum environment of $5 \times 10^{-5}$ Pa for 1 h to release residual stress.

The final $Ti_3SiC_2$ bulk samples were irradiated with 110 keV He$^+$ beam incident at 0° to the normal using the tandem accelerator at Institute of Applied Physics, Chinese Academy of Science [25]. The typical irradiation flux was kept at ~$7.0 \times 10^{11}$ ions·cm$^{-2}$·s$^{-2}$. The irradiation fluence delivered to the samples was $5 \times 10^{16}$ ions·cm$^{-2}$ and the background pressure during irradiation was $< 5 \times 10^{-4}$ Pa. The total damage, measured in terms of dpa, was simulated using

SRIM-2013 [26], with displacement energies of 25, 15, and 28 eV for Ti, Si, and C, respectively, and with an average atomic density for $Ti_3SiC_2$ of $8.34\times10^{22}$ atoms·cm$^{-3}$. The damage level obtained from the SRIM-2013 simulation was estimated to be 0.5 dpa at the surface, rising to 2.8 dpa at a depth of 400 nm as shown in Fig. 1.

GIXRD data were obtained at beam line BL14B1 of the Shanghai Synchrotron Radiation Facility at a wavelength of 1.2398 Å. The size of the focus spot was ~0.5 mm and the end station was equipped with a Huber 5021 diffractometer. The diffraction data were analyzed using a Rietveld analysis program (Rietica 7.1). From the Rietveld refinement, the $Ti_3SiC_2$ phase was clearly identified from the sharp peaks at the relevant diffraction angles. The damaged component was identified by broader and slightly displaced peaks from the $Ti_3SiC_2$ peaks. The uncertainties in the phase compositions and lattice parameter were typically 1-2% and 0.5%, respectively.

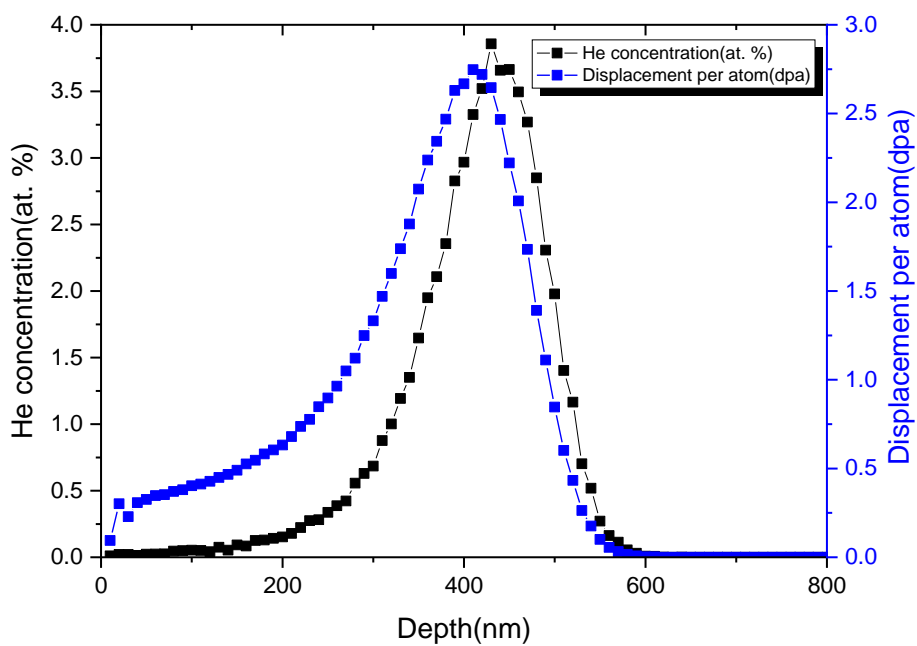

*Fig. 1* He atom concentration and radiation damage in dpa for $Ti_3SiC_2$ irradiated with $5\times10^{16}$cm$^{-2}$ simulated by SRIM 2013.

Raman spectroscopy was performed on a XploRA Laser Raman spectrometer produced by HORIBA Jobin Yvon. The measurements were conducted using a 632.8 nm wave-length laser with a detection range from 100 to 1900 cm$^{-1}$ and a total acquisition time of 100 s. The spectrum resolution was better than 2 cm$^{-1}$. The Raman signal is known to decay exponentially as a function of the distance from the surface, with a decay length of approximately 10 nm. The Raman spectroscopy should, therefore, be sensitive to the surface region.

The hardness and modulus of the samples were measured using Hysitron TI-950 Tribo Indenter Nanoindenter and Atomic Force Microscope (AFM) at room temperature with a nano indenter by the continuous stiffness measurement (CSM) mode equipped with a diamond Berkovich indenter (triangular based pyramid). The depth of the indentation is about 500 nm. 10 indentations were performed for each sample and the results were averaged over the ten measurements. The real shape of the indenter was calibrated by the standard method that involves indenting a fused silica sample at different normal loads.

In order to characterize the evolution of the He bubbles, the topography, and the microstructural evolution of the samples, TEM observations were carried out using a FEI Tecnai G2 F30 transmission electron microscope, in the Institute of Metal Research, Chinese Academy of Science. Cross-sectional TEM samples were prepared using mechanical polishing and then ion milling to form a wedge to create sufficient electron transparency.

## 3. Results

Structural changes and lattice swelling after irradiation were characterized by GIXRD and refinement. Fig.2 shows the GIXRD patterns and the refinement results for the unirradiated Ti$_3$SiC$_2$ sample and from samples irradiated at RT and at 750 °C . X-ray incident angle was 1.5°, corresponding to the penetration depth of 457 nm, which is close to the helium irradiation depth range.

Compared to the unirradiated sample's pattern, the pattern for the RT sample shows a clear increase in the background signal, a significant decrease in the peak intensity, and a notable broadening of the peaks. These changes suggest a decrease in the crystallinity of the lattice. There is also a small shift of the peak positions to the low angle direction, indicating an expansion in the unit cell after helium irradiation. The refinement results showed that the $c$ lattice parameter increased from 17.65 to 17.95 Å, which was an increase of approximately 1.70 %. On the other hand, the $a$ lattice parameter remained almost unchanged at 3.06 Å. This is in good agreement with previous RT-irradiation result [18], which showed an increased $c$ lattice parameter of 17.9 Å and an almost unchanged $a$ lattice parameter of 3.06 Å. It is possible that the increase of the $c$ lattice parameter mostly comes from the lattice expansion induced by the He bubbles and clusters.

The 750 °C irradiated sample exhibits a much higher crystallinity level compared with the RT one. The pattern is almost identical to that of the unirradiated sample. The $c$ lattice parameter increases slightly from 17.65 to 17.70 Å, which is only by 0.3%. In the previous study, where $Ti_3SiC_2$ was irradiated by He ion at 450 °C [19], Rietveld refinement showed a $c$ lattice parameter of 17.81 Å. Therefore, the swelling along the $c$ direction at 750 °C is smaller than that at 450 °C. The results indicate that $Ti_3SiC_2$ has a very good tolerance to swelling induced by He irradiation at a higher temperature of 750 °C, underscoring the potential of $Ti_3SiC_2$ as cladding and structure materials at this temperature.

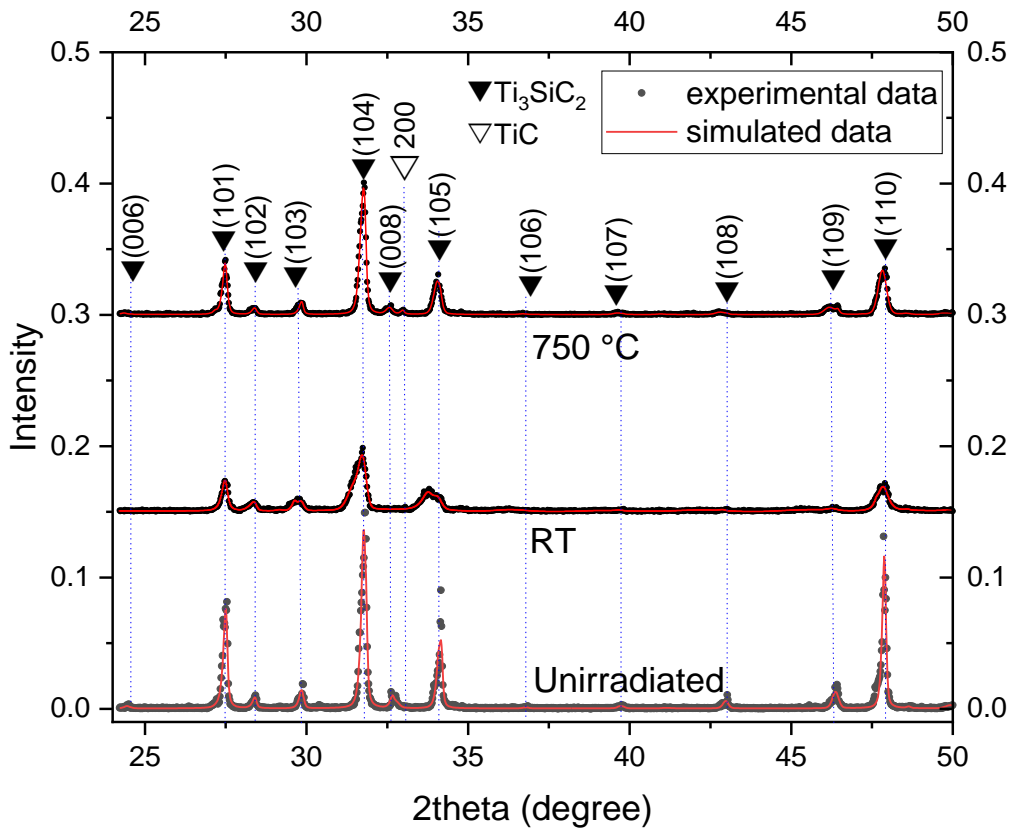

*Fig.2 GIXRD and refinements results for samples unirradiated, irradiated at RT and 750 °C. One phase ($Ti_3SiC_2$ undamaged phase) was used to refine the spectrum of unirradiated $Ti_3SiC_2$. Three phases were used to do the refinement of the spectra of RT and 750 °C irradiated samples:*

*(i) TiC phase (to represent TiC or fcc-(Ti$_3$Si)C$_2$), (ii) Ti$_3$SiC$_2$ undamaged phase, and (iii) Ti$_3$SiC$_2$ damaged phase. The red lined spectra are the simulated spectra based on the experimental data.*

The near surface damage was analyzed by Raman spectroscopy. Spectra collected for the unirradiated and irradiated Ti$_3$SiC$_2$ samples are shown in Fig. 3. The spectrum of the unirradiated sample has six peaks at 159, 228, 281, 312, 631, and 678 cm$^{-1}$, which corresponds to Ti$_3$SiC$_2$ [27,28]. After helium irradiation at RT, all these peaks disappear. Two notable peaks are located at ~1335 and 1580 cm$^{-1}$ in the spectrum, which are associated with the A$_{1g}$ and the E$_{2g}$ vibrational modes of graphite. These results are in good agreement with previous studies, which reported disappearance of Ti$_3$SiC$_2$ related peaks and appearance of graphite related peaks [19,28]. These changes imply a significant surface damage. Two small TiC-related peaks located at 386 and 590 cm$^{-1}$ (shown as red dashed lines) are found in the Raman spectra, suggesting that there is a small amount of TiC$_x$ in the near surface region. In contrast, for the 750 °C irradiated Ti$_3$SiC$_2$, the spectrum is almost the same as that for the unirradiated sample, and all peaks corresponding to Ti$_3$SiC$_2$ are still present in the spectrum. Peaks associated with the A$_{1g}$ and the E$_{2g}$ vibrational modes of graphite are also visible in the 750 °C irradiated Ti$_3$SiC$_2$ spectrum. These two peaks were possibly caused by carbon contamination in the background vacuum system during helium irradiation. Since the Raman spectroscopy analysis only detects the state of a surface region, the results indicate that the surface region exhibits good crystallinity at an irradiation temperature of 750 °C, and that Ti$_3$SiC$_2$ has much better tolerance to surface damage for irradiation at 750 °C than at RT.

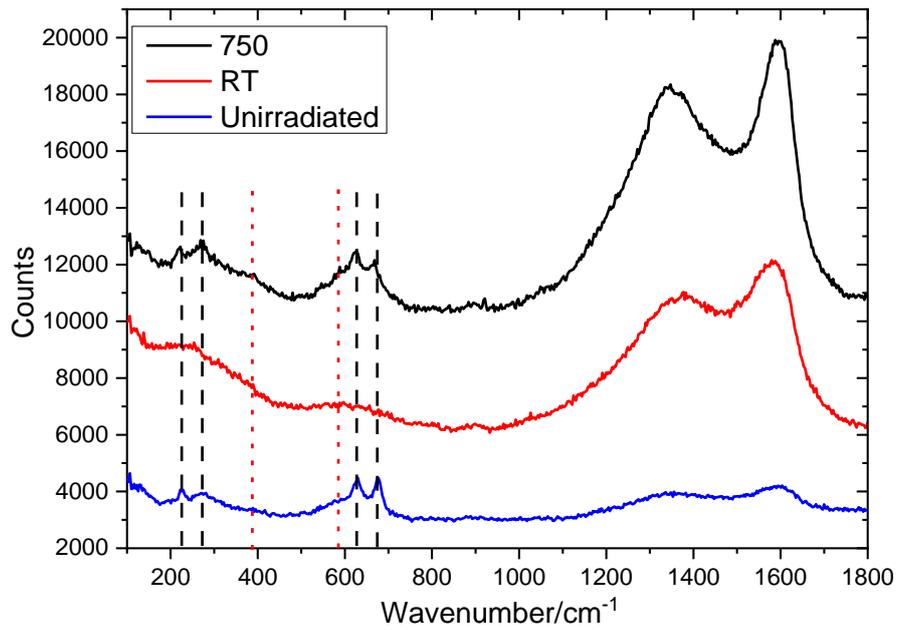

***Fig.3*** *Raman spectra for unirradiated sample, samples irradiated at RT and 750 ℃*

Hardness and Young's modulus before and after irradiation were measured by nano indentation and are shown in Fig. 4. The hardness of $Ti_3SiC_2$ changes from 6.68 to 9.96 GPa after RT irradiation, with a significant increase of 49.1%. However, when irradiated at 750°C with the same irradiation fluence, the hardness only changes from 6.68 to 7.21 GPa, with an increase of 7.93%. In contrast to hardness, Young's modulus decreases due to irradiation. For the RT irradiated sample, the Young's modulus drops from 186.51 to 130.11GPa, with a decrease of >30.2%. For the 750°C irradiated sample, Young's modulus drops from 186.51 to 167.65 GPa, with a decrease of ~10.1%. For the 750 °C irradiated sample, the changes in the hardness and Young's modulus are much smaller than those for the RT irradiated sample, which confirms that the $Ti_3SiC_2$ has better tolerance for irradiation induced mechanical properties change at 750 °C than that at RT, although changes of ~8-10% in mechanical properties observed at 750 °C are not insignificant.

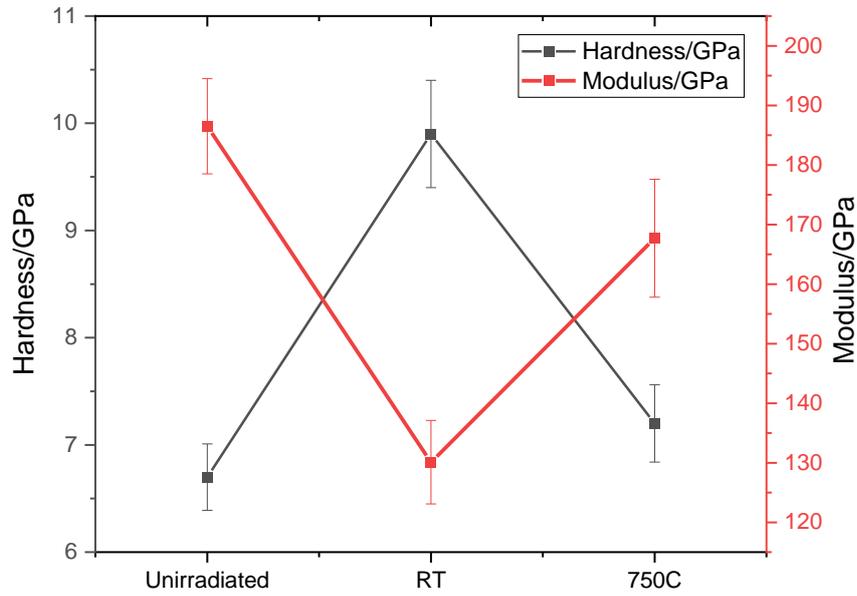

*Fig.4* Hardness and Young's modulus of the unirradiated sample and the samples irradiated at RT and 750 ℃

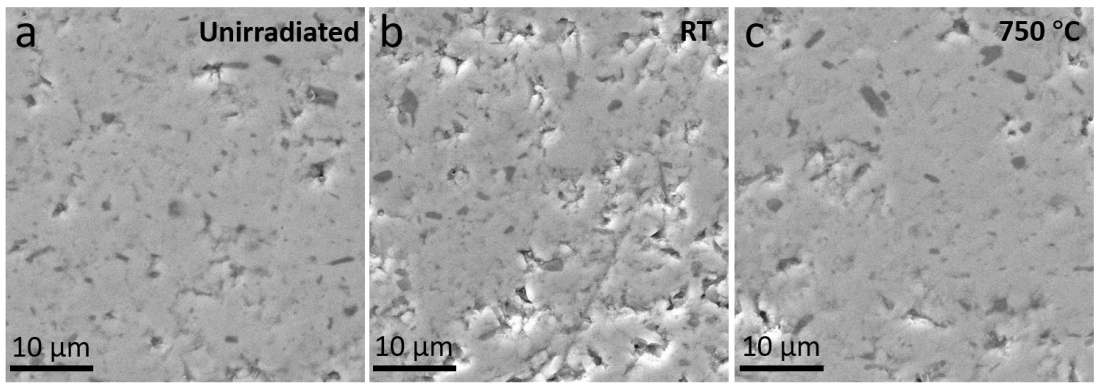

*Fig. 5* SEM images of (a) unirradiated $Ti_3SiC_2$, and $Ti_3SiC_2$ irradiated at (b) RT, and (c) 750 ℃.

Fig. 5 shows the SEM images of the unirradiated and irradiated $Ti_3SiC_2$ samples. The darker spots are $Al_3O_2$, which was introduced by Al doping during the sample fabrication using the reactive sintering method. Unlike $Ti_3AlC_2$ and $Ti_2AlC$, which show a high density of cracks after RT irradiation [9,23,24], the surface of $Ti_3SiC_2$ is free of cracks at both RT and 750 ℃, suggesting that $Ti_3SiC_2$ has a better resistance to irradiation-induced surface cracks.

In order to analyze the morphology and the distribution of He bubbles as well as to analyze the irradiated $Ti_3SiC_2$ in TEM, bright field TEM images of the $Ti_3SiC_2$ after irradiation at RT and at 750°C from [11-20] are shown in Figs. 6a and 6b, respectively. In the RT-irradiated sample, most of the helium bubbles are spherical in shape, with diameters of ~1-2 nm. The He bubbles are densely distributed in the irradiated area. In the 750°C-irradiated sample, He bubbles exhibit very different shape; most of them are columnar with a very long axis and a small cross section. More specifically, the diameter for most of the helium bubbles is <1 nm, whereas the length is >15 nm, with some bubbles as long as 40 nm. Moreover, the amount of He bubbles was found to increase monotonically with the dpa. The density of He bubbles is the highest in the damage peak region and the lowest in the near-surface region. In addition, as shown in Fig. 6c, helium bubbles are distributed parallel to the nano-layer direction. The results indicate the structure of $Ti_3SiC_2$ has a great ability to impede the free growth of the He bubbles at elevated temperatures and therefore has a high resistance to irradiation-induced swelling during the He irradiation and He evolution. Fig. 6d shows the morphology and distribution of helium bubbles at a grain boundary in the 750 °C-irradiated sample. Density of the helium bubbles is evidently higher at the grain boundary than in the grain interiors. The helium bubbles still exhibit a columnar shape at the grain boundary. There is an obvious depleted zone of ~100 nm between the He bubbles in the GB region and in the bulk region. According to our TEM observations, when the dose is larger than 0.5 dpa, the effects of the irradiation depths and of the specific value of dpa on the shape of He bubbles are very small. The main factor that controls the shape of He bubbles is the irradiation temperature. When the temperature reaches 750 °C, the shape of the He bubbles will be elongated. These results show that the He bubbles preferentially concentrate at defects such as grain boundaries.

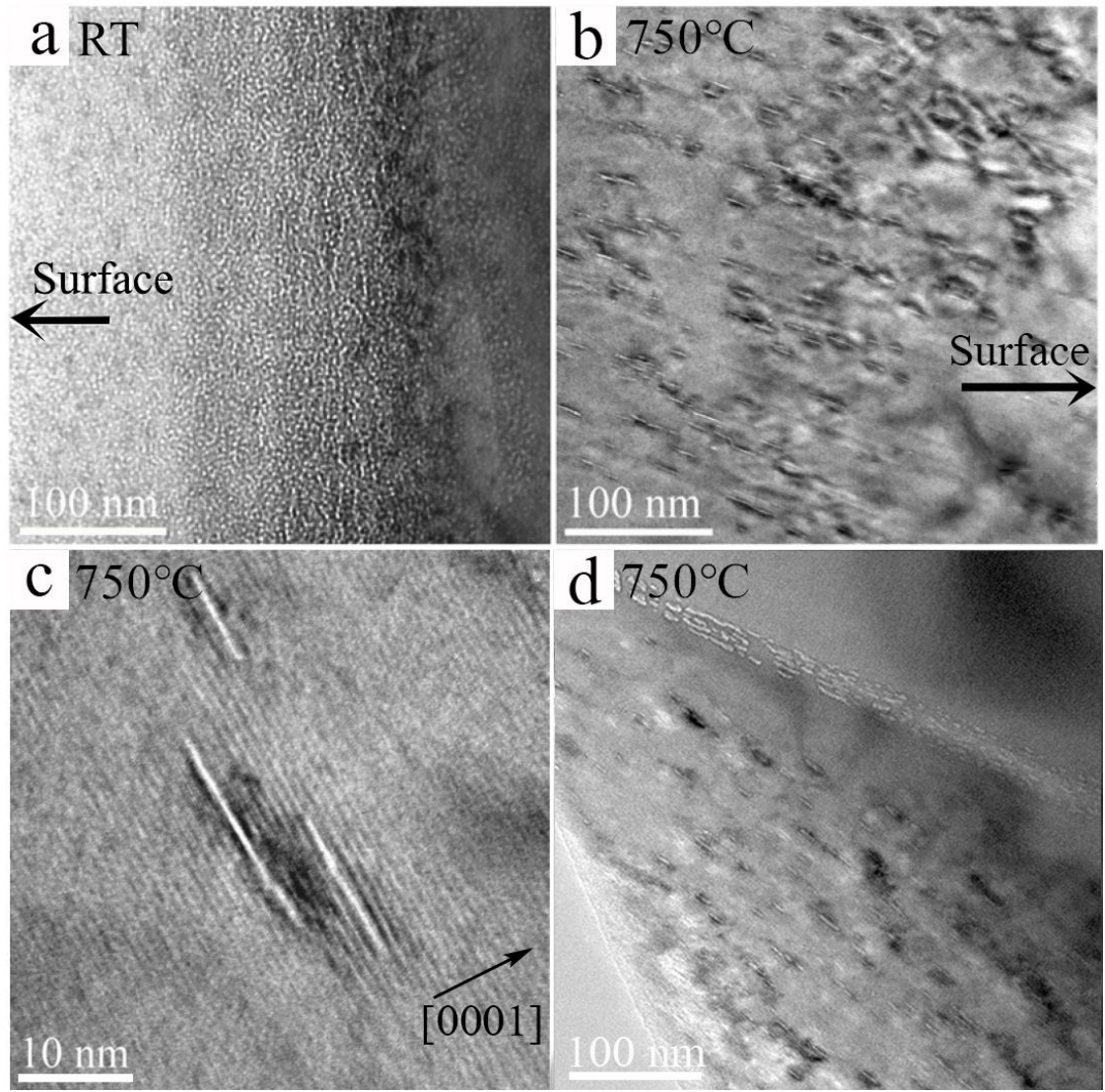

***Fig.6*** *TEM image of the Ti$_3$SiC$_2$ samples, observed from [11-20], irradiated at (a) RT, the white spots are He bubbles, the black region is the damage peak region which has a high density of black spots defects and (b) 750 ℃. (c) HRTEM of the Ti$_3$SiC$_2$ samples irradiated at 750 ℃ showing a He bubble located in the bulk and aligned parallel to basal plane. (d) TEM image of the Ti$_3$SiC$_2$ samples irradiated at 750 ℃ with a grain boundary, showing a zone depleted in He bubbles.*

## 4. Discussion

### 4.1 Irradiation induced hardening

The hardness of the Ti$_3$SiC$_2$ irradiated at RT and at 750°C increased compared with that of the unirradiated sample. The hardness change for the Ti$_3$SiC$_2$ irradiated at 750°C was smaller than that for the RT irradiation. A possible mechanism underlying the radiation-induced hardening is dispersed barrier hardening [29], where radiation induced defects (such as vacancies or interstitial clusters) act as barriers to movement of dislocations. For the He irradiated $Ti_3SiC_2$, the irradiation-induced hardening could come from helium bubbles as well as the He atoms and clusters.

For the He bubbles, the effect of the bubble formation on hardness can be captured by the following relation [30,31]:

$$\Delta H = 3\Delta\sigma_{cavity} = 1/8 \ MGbdN^{2/3}$$

where $\Delta H$ is the change of the hardness due to irradiation, $\Delta\sigma_{cavity}$ is the change in strength induced by bubbles, $M$ is the Taylor factor reflecting crystal orientation, $G$ is the shear modulus (GPa), $b$ is the length of the Burgers vector (nm) of the dislocation, $d$ is the cavity diameter (nm), and $N$ is the cavity density ($m^{-3}$). The density of the helium bubbles was determined by counting the number of He bubbles at the same depth in the RT and 750 °C irradiated samples in an area of 50nm × 50 nm and then dividing the number of bubbles by the product of area and assumed thickness (80 nm). The density of the helium bubbles in the 750 °C and RT irradiated samples are significantly different from each other and they are $\sim 5.0 \times 10^{24}$ $m^{-3}$ for RT-irradiated sample (with an average size of 1 nm) and $\sim 1.0 \times 10^{23}$ $m^{-3}$ for the 750 °C-irradiated sample (with an average size of 15 nm in the longer direction; since He bubbles are elongated, 15nm is the upper limit and a significant overestimate of the bubble size). According to the equation above, the product of $d \cdot N^{2/3}$ is much higher in the RT-irradiated sample than in the 750 °C-irradiated sample, which can explain a more significant hardness increase. This is one of the reasons for the much higher irradiation-induced hardening in $Ti_3SiC_2$ irradiated at RT.

Although many helium atoms bind to vacancies to form helium bubbles in the irradiated area, there are still large number of isolated helium atoms or helium clusters in the RT-irradiated sample [26-27]. A fluence of $5 \times 10^{16}$ $cm^{-2}$ He into a 100 nm depth range corresponds to a He density of $5 \times 10^{27}$ $m^{-3}$ at the peak region. The number density of the He bubbles in this region can be estimated using the relationship between the He bubble pressure and the number of He atoms [34]:

$$P = 4.83 \times 10^2 \exp(5.15 \times 10^{-23} \rho) \text{ atm}$$

where $P$ is the pressure of the helium bubble and $\rho$ is the He bubble density in the units of He atoms/$m^3$. Assuming a bubble pressure of 10 GPa [35,36], we can estimate that each 1.0 nm diameter bubble contains approximately 500 He atoms, which predicts the bubble density of 1 ×$10^{25}$ m$^{-3}$ while the actual He bubble density determined from the TEM images is only ~5× $10^{24}$ m$^{-3}$. This analysis indicates that in addition to the He bubbles, there are large number of isolated helium atoms and helium clusters in the RT-irradiated sample. Previous studies have shown that high concentrations of helium interstitials can lead to noticeable hardening, especially when the helium irradiation dose reaches the dpa higher than 1.0 [29-30]. The large amount of isolated helium atoms and clusters are another reason for the higher hardness in the RT irradiated Ti$_3$SiC$_2$.

**4.2 The relationship between irradiation temperature and characteristics of helium bubbles**

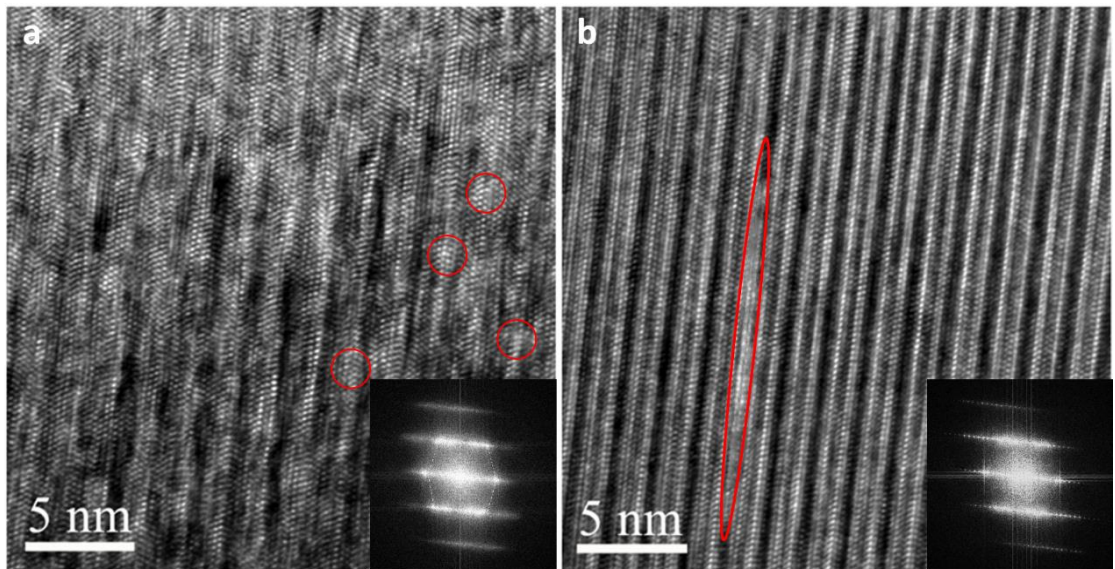

*Fig.7* HRTEM and the corresponded fast Fourier transform (FFT) patterns of the Ti$_3$SiC$_2$ samples, viewed from [11-20], irradiated at (a) RT showing the disorder of the layer structure in the He bubbles(red circles) region and (b) at 750 °C showing a He bubble(red oval) located between the layers while the layer structure remained intact.

In the RT irradiated sample, the helium bubbles had a spherical shape, a size of ~1 nm, and they were randomly distributed. In addition, the nano-laminate layered structure of Ti$_3$SiC$_2$ was found to be severely damaged, as shown in Fig. 7a. The FFT pattern also indicates a significantly damage because of the distortion and disappearance of some diffraction spots, as well as the hexagonal outline in the center of the FFT pattern (marked in the dashed lines). The formation of the small-sized, spherical He bubbles at RT is mainly due to the He implantation and redistribution. The helium redistribution at RT is mostly driven by cascade-collision-induced atomic displacements [37]. These displacements are randomly located in the damage peak region, and the displacement range is usually very small. Consequently, the He redistribution distance is very limited. As a result, He bubbles can only absorb helium atoms and clusters located very close to them and grow into small bubbles (here with diameters of only ~1 nm). These He bubbles are randomly distributed between Ti, Si, and C planes. First principle calculations of single He atom in $Ti_3SiC_2$ have shown that at higher temperatures (>500 °C) He atoms will quickly migrate into the Si layer whereas at RT some of them will be trapped by vacancies in the C layer and Ti layer [39]. Thus, at RT, the trapped He atoms in C and Ti layers may not have enough energy to detrap and move to the Si layer. Instead, they can bind nearby He atoms, forming He bubbles.

In the 750°C irradiated sample, as mentioned before, He bubbles are much more elongated and have a lower density. In addition, these helium bubbles are located between the layers and are oriented parallel to the nano-laminate layer of the $Ti_3SiC_2$ while the layer-structure remains intact (Fig. 6c and Fig. 7b). The FFT pattern showing clear diffraction spots indicates the higher preservation of the crystallinity than that at RT. There are several reasons for the growth and the evident shape change of the helium bubbles at this elevated temperature. First, a previous study has shown that the formation energy of a He interstitial in the Si layer is 2.99 eV, which is lower than the He interstitial formation energy in the C layer (3.11 eV) and in the Ti layer (5.10 eV) [40], which means that He atoms prefer to stay in Si planes. In addition, the diffusion barrier for He atoms in the Si layer is only 1.17 eV (and it is the lowest among the different layers), He atoms can migrate in the Si layer at the temperatures above 500 °C [39]. These calculations have also indicated that He atoms in other layers will quickly migrate into the Si layer at higher temperatures (>500 °C) [39,41,42]. Therefore, at 750 °C, the He atoms and clusters that might have formed in the Ti and C layers, will migrate to the Si layers and then diffuse within this layer [37] until they are finally retrapped by another He bubble or vacancy. Hence, at 750 °C, a large fraction of helium atoms, clusters and bubbles gather in this layer, and can form the extremely elongated helium bubbles.

## Conclusions

Helium irradiation of $Ti_3SiC_2$ at RT and 750°C was conducted to compare the tolerance of this material to irradiation damage at different temperatures. Significant increase of the resistance to He irradiation induced swelling and hardening was observed at 750 °C. The bubbles were located in the nano-laminate layers of $Ti_3SiC_2$, whereas the overall crystal structure remained intact at 750 °C. The ability to limit the free growth of He bubbles into larger-sized-spherical shape and the smaller changes of mechanical properties at the higher temperature indicate that $Ti_3SiC_2$ has a good resistance to He effects at high temperatures and that its unique structure can accommodate large amounts of helium in the nano-laminate layer, with its layered structure remaining unchanged.

## Acknowledgements

I. Szlufarska and H. Zhang acknowledge financial support from the US Department of Energy, Basic Energy Science Program under Award # DEFG02-08ER46493. The authors thank beam line BL14B1 (Shanghai Synchrotron Radiation Facility) for providing the beam time. The authors acknowledge the financial support from China Scholarship Council and the National Nature Science Foundations of China under grant number 11375046 and U1630107.